\documentclass[aps, prl, reprint, showpacs, twocolumn, 10pt, groupedaddress]{revtex4-1}

\usepackage{graphicx, amsmath, dsfont, dcolumn, bm, times, color, braket, amssymb, multirow}

\begin{document}

\bibliographystyle{apsrev4-1}

\title{Braiding and all quantum operations with Majorana modes in 1D}

\author{Viktoriia Kornich}
\affiliation{Kavli Institute of Nanoscience, Delft University of Technology, 2628 CJ Delft, The Netherlands}
\author{Xiaoli Huang}
\affiliation{Kavli Institute of Nanoscience, Delft University of Technology, 2628 CJ Delft, The Netherlands}
\author{Evgeny Repin}
\affiliation{Kavli Institute of Nanoscience, Delft University of Technology, 2628 CJ Delft, The Netherlands}
\author{Yuli V. Nazarov}
\affiliation{Kavli Institute of Nanoscience, Delft University of Technology, 2628 CJ Delft, The Netherlands}

\date{\today}

\begin{abstract}
We propose a scheme to perform braiding and all other unitary operations with Majorana modes in 1D that, in contrast to previous proposals, is solely based on resonant manipulation involving the first excited state extended over the modes. The detection of the population of the excited state also enables initialization and read-out. We provide an elaborated illustration of the scheme with a concrete device.
\end{abstract}

\maketitle

\let\oldvec\vec
\renewcommand{\vec}[1]{\ensuremath{\boldsymbol{#1}}}

The paradigm of topological quantum computation \cite{kitaev:anyons, nayak:rmp08} provides an elegant solution to the most important problem in quantum manipulations: decoherence problem. It implements a topologically protected degenerate ground state as a computational basis. The degenerate state can be visualized as a set of localized anyons while unitary operations are performed by adiabatic exchange of the anyons, that is, braiding of their world lines \cite{nayak:rmp08}. The braiding is feasible in 2D and impossible in 1D since anyons should not collide in the course of operation. The intrinsically slow speed of adiabatic manipulation, as well as the difficulties of read-out and initialization of the protected states, should be compensated by the intrinsic fault-tolerance of the operations.

Of all numerous physical realizations of topologically protected degenerate ground state proposed, the Majorana zero-energy states in hybrid semiconductor-superconductor devices\cite{lutchyn:prl10, oreg:prl10} seems to be the most technologically advanced and elaborated. After pioneering experiments \cite{mourik:science12}, an enormous outgoing research effort \cite{ deng:nanolett12, das:natphys12, finck:prl13} resulted in considerable improvement of the technology and new observations, yet the quantum coherence in degenerate subspace still awaits experimental demonstration \cite{WimmerReview2020}. An obvious difficulty is that Majorana modes are realized in 1D nanowires, making direct braiding impossible. In principle, the 1D wires can be combined into a 2D network. There are elaborated schemes to realize braiding in various systems, for instance, in T- or Y-junctions of nanowires, \cite{alicea:natphys11, harper:prresearch19, yang:prb19, posske:prr20, aasen:prx16, clarke:prb17, beenakker:scipost20}. A enormous technological challenge to make such networks with necessary controls is being addressed \cite{networks}, but the progress is slow so far. 



In this Letter, we propose a scheme to realize Majorana braiding in a single 1D nanowire. Eventually, with this scheme one can realize any unitary transformation in the degenerate subspace, as well as initialization and read-out in this subspace. The scheme uses resonant manipulation technique, the resonance being between the degenerate subspace and the lowest excited state that extends over all Majorana modes. The initialization and read-out is possible if the population of the excited state is detected. 

Strictly speaking, the scheme compromises the quantum computation paradigm since the topological protection fails during the operation. The system is subject to relaxation while being in the excited state. There are standard means to reduce this only source of decoherence, for instance, photonic \cite{yablonovitch:prl87, zheludev:natm12, blais:pra04} and phononic \cite{pennec:surfsr10, deymier:book13, khelif:book15, laude:book15} cavities and metamaterials, and make the operation time shorter than the corresponding relaxation time. It is important that the protection is preserved between the operations. This makes the scheme an ideal tool to demonstrate persistence of quantum superpositions in the degenerate subspace, and quantify the macroscopically long decoherence time expected. In the final part of the Letter, we discuss the use of the scheme in wider context. We illustrate the scheme on the example of a minimum concrete setup, at general level as well as with a concrete microscopic model and numbers.

\begin{figure}[tb]
\begin{center}
\includegraphics[width=0.92\linewidth]{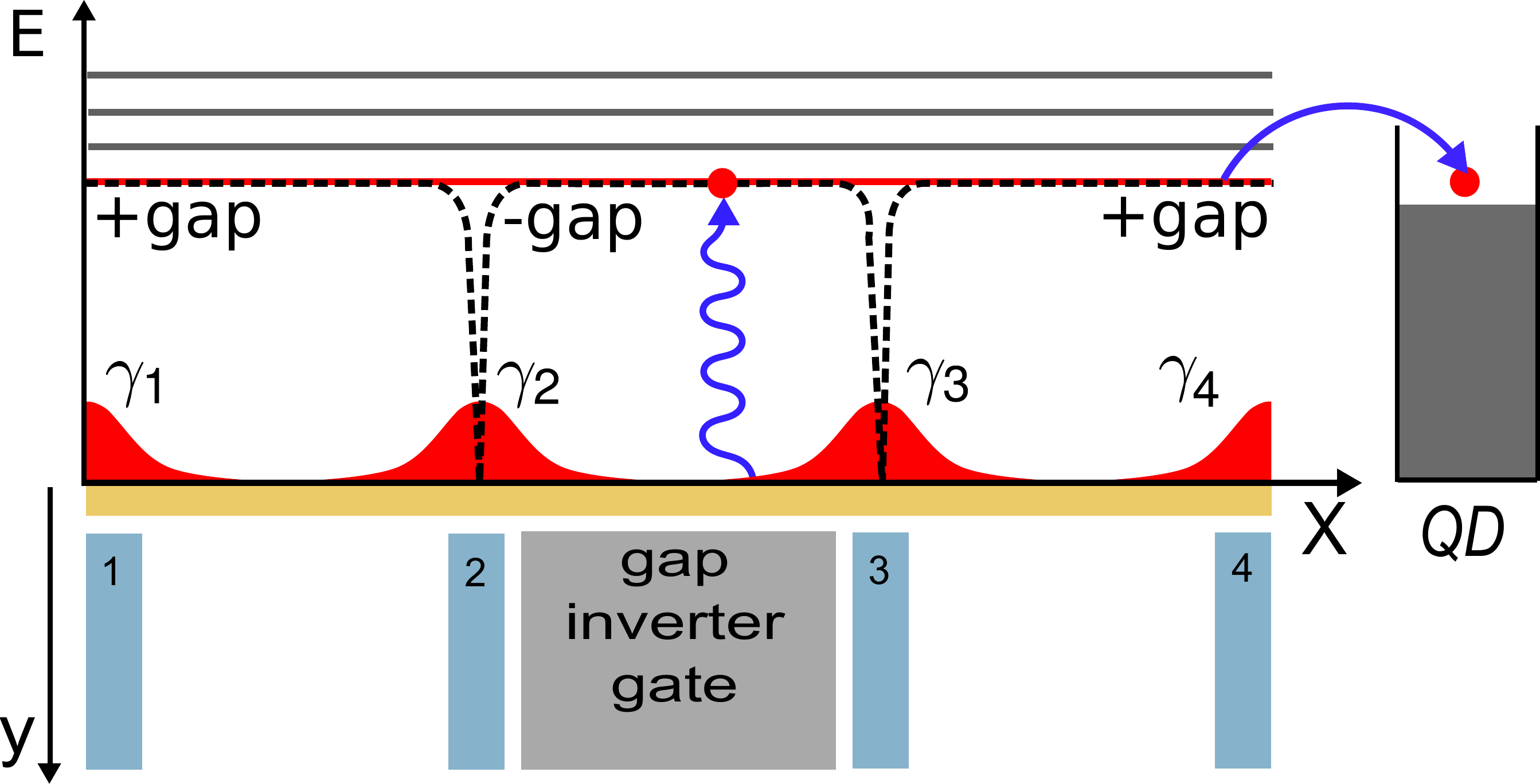}
\caption{The setup for the resonant manipulation of Majorana modes. The proximitized nanowire (orange rectangle) with the inverted gap in the middle section hosts four Majorana modes ($\gamma_{1-4}$) formed on the edges of the sections of different topology. The four gates in the vicinity of the modes are used to apply a pulse sequence for the resonant manipulation, the resonance being with the lowest excited state (red line) extented over the modes. A quantum dot on the left can be used to detect the population of the excited state. }
\label{fig:setup}
\end{center}
\end{figure}

The setup under consideration (Fig. \ref{fig:setup}) encompasses a finite 1D wire brought in proximity with a superconductor. It hosts 4 localized Majorana modes, two at the ends and two in the middle. This is achieved by a gap inversion in the middle section of the wire by a nearby gate.  The wire sections at the sides are thus in topological regime of parameters while the middle section is topologically trivial. It is important for us that the first excited state right above the gap extends over the whole wire. This is achieved by matching the absolute values of the gap in the middle and side sections by the gap inverter gate. To achieve efficient resonant manipulation, we require four more gates near the positions of Majorana modes. This is all we need for resonant manipulation. To detect a possible quasiparticle in the excited state, we put a quantum dot nearby (it can be in the same nanowire, as presented in \cite{deng:prb18, prada:prb17}). The addition energy of the dot is tuned such that a quasiparticle in the excited state tunnels to the dot changing its charge, which is measured. For effective detection, the tunnel rate should exceed the relaxation rate. The tunnel coupling can be switched on only for duration of measurement. 

To start with, let us understand the basis involving the Majorana modes and the first excited state. Let $c_L=(\gamma_1+i\gamma_2)/2$, $c_R=(\gamma_3+i\gamma_4)/2$ be the quasiparticle annihilation operators in Majorana subspace, and $c_{ex}$ to be that in the excited state. A basis state is defined as $|n_{L}, n_{R}\rangle|n_{ex}\rangle$, where $n_{L}, n_{R}, n_{ex} = 0,1 $ are the respective occupation numbers. We thus have 8 states. They separate into two groups of four corresponding to two possible total parities. There can be no coherence between the states of different parities. We define the bases as follows:
\begin{eqnarray}
\label{eq:basiseven}
\Phi_e=\{|00\rangle|0\rangle,\ \ |11\rangle|0\rangle,\ \ |01\rangle|1\rangle,\ \ |10\rangle|1\rangle\},
\end{eqnarray}
for the even parity, and
\begin{eqnarray}
\label{eq:basisodd}
\Phi_o=\{|01\rangle|0\rangle,\ \  |10\rangle|0\rangle,\ \ |00\rangle|1\rangle,\ \ |11\rangle|1\rangle\},
\end{eqnarray}
for the odd parity. The first two states for each parity form Majorana subspace.
We can thus realize a Majorana qubit for each parity. We would like to perform unitary operations in Majorana subspace. A particular unitary operation is a braiding of two Majorana modes defined as $U_{ij}=\frac{1}{\sqrt{2}}(1+\gamma_i\gamma_j)$. For instance, the braiding of the second and the third mode in the odd basis $\Phi_e$ is given by
\begin{eqnarray}
\label{eq:Braiding}
U_{23}^o=\frac{1}{\sqrt{2}}(1+\gamma_2\gamma_3)=\frac{1}{\sqrt{2}}\begin{pmatrix}1 & i & 0 & 0\\
i & 1 & 0 & 0\\
0 & 0 & 1 & i \\
0 & 0 & i & 1\end{pmatrix}.
\end{eqnarray} 
As we see, it is separated into blocks of Majorana and excited subspace, as these operations are independent. Since we wish to operate in Majorana subspace, the excited block is irrelevant. The corresponding matrix in the even subspace is obtained from (\ref{eq:Braiding}) by the following transformation
\begin{eqnarray}
\label{eq:UeUo}
U^e=\Sigma_y\sigma_yU^{o*}\sigma_y\Sigma_y.
\end{eqnarray} 
$\sigma_y,\Sigma_y$ being Pauli matricies acting within and over the blocks, respectively. Eventually, this relation holds for all braidings as well as for any $4 \times 4$ matrix we consider here. So we wish to perform braidings, as well as any unitary operations in Majorana subspace. This task by its own is senseless unless we have means to initialize to a state in this subspace and measure the result. Let us see how we can realize this by resonant manipulation.

A resonant manipulation is performed by applying the oscillating voltages to the gates $1-4$ with the frequency matching the energy spacing. At constant amplitudes, the general Hamiltonian in rotating wave approximation reads: 
\begin{eqnarray}
\label{eq:Hrm}
H_{\rm rm}=\left(\alpha_1 c_L+\alpha_2 c_R +\alpha_3 c_L^\dagger +\alpha_4 c_R^\dagger\right) c_{\rm ex} + h.c. 
\end{eqnarray}
  The four complex coefficients $\alpha_{1-4}$, are in linear relation with the four complex voltage amplitudes at the gates, so 4 gates suffice to control all coefficients. Applying a pulse of duration $t$ makes a unitary operation $U=e^{-iH_{\rm rm}t}$ in 8-dimensional basis. The manipulation conserves parity, so the matrix separates in two $4 \times 4$ blocks $U_e, U_{o}$ in the bases $\Phi_e, \Phi_{o}$. It is simple and important to show that these matrices satisfy the same relation (\ref{eq:UeUo}) as the braiding matrices. 
  
Let us stress that our aim is to find a unitary transformation that works in Majorana subspace only. To this end, we require a special form of the resulting $U$: that separated in two $2 \times 2$ blocks, like in Eq. \ref{eq:Braiding}. In other words, the excited state should not be populated at the end of the resonant manipulation if we start in Majorana subspace. This is impossible to achieve with a single pulse. A key observation is that this can be achieved combining {\it several} pulses. Two pulses with 8 complex parameters in total in principle suffice to realize our aim: an arbitrary $2 \times 2$ unitary transformation in Majorana basis. We describe the concrete methods of the pulse design and give examples further in the text.

\begin{figure}
	\centerline{\includegraphics[width=0.48\textwidth]{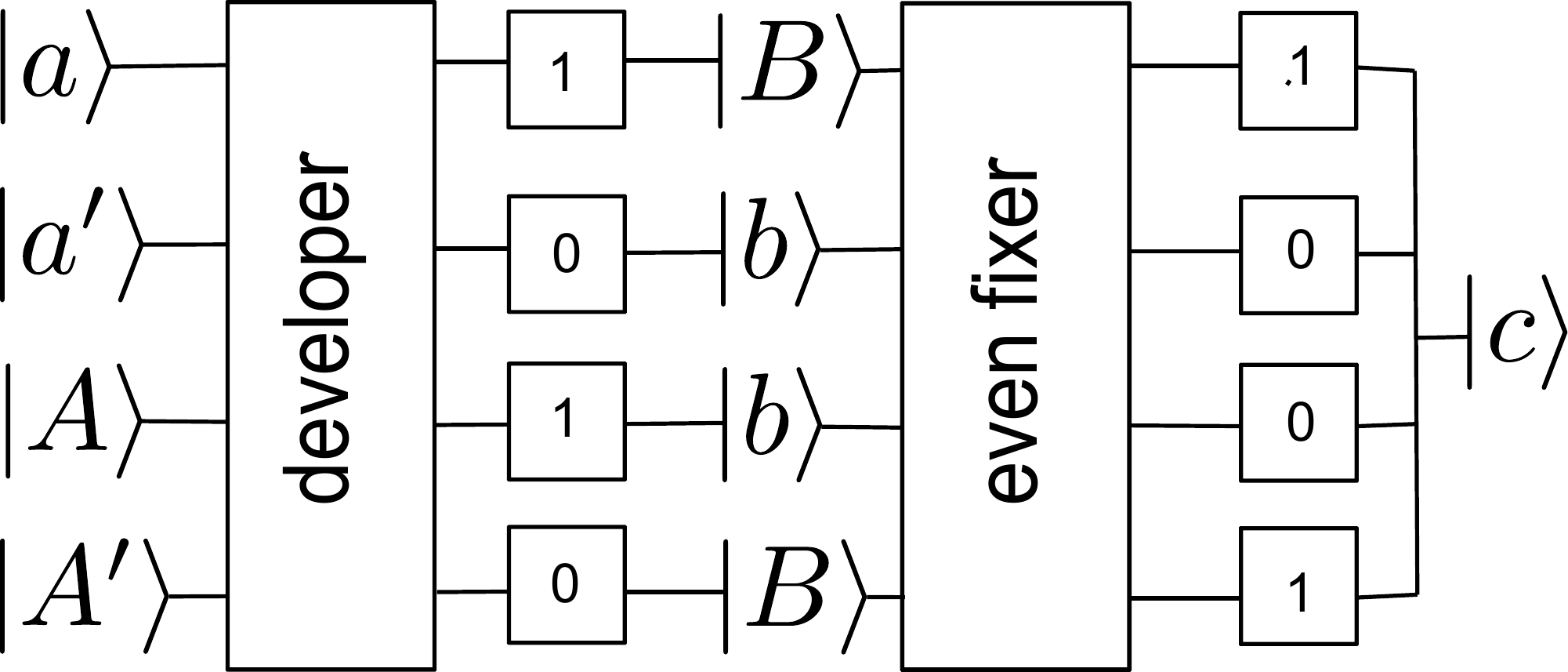}
	}
	\caption{Initialization and read-out in Majorana subspace is achieved by two resonant pulse sequences: developer and fixer, and subsequent measurements of the excited quasiparticle. Upper(lower)case letters refer to Majorana superpositions of odd(even)parity, prime indicates orthogonality, $\langle a'|a\rangle = 0$. The measurement outcomes are in square boxes. The protocol brings the system to the state $|c\rangle$ from an unknown state. The probabilities of the measurement outcomes give the probabilities of the states $|a\rangle,|a'\rangle,|A\rangle,|A'\rangle$. See the text for details. }
	\label{fig:developing}
\end{figure}
Let us describe the protocol for initialization and read out starting from an unknown state of unknown parity in Majorana subspace. We will show that this requires two resonant pulse sequences, that is, unitary transformations, and a measurement after each sequence. We dub these sequences a developer and a fixer. To start with, let us assume that we start in a Majorana state of even parity. Let us understand the effect of the following $4 \times 4$ unitary transformation:
\begin{eqnarray}
\label{eq:pulse1e}
D^{e}=|B\rangle|1\rangle\langle a|\langle0|+ |b\rangle|0\rangle\langle a'|\langle0|
\\ \nonumber
+|B'\rangle|1\rangle\langle A|\langle1|-|b'\rangle|0\rangle\langle A'|\langle1|.
\end{eqnarray}
Here, lowercase letters denote the Majorana states in the even subspace ($|00\rangle$, $|11\rangle$ or their linear combination) while capital ones denote  those in the odd subspace ($|01\rangle$, $|10\rangle$ or their linear combination). The prime denotes a corresponding orthogonal state, $|a'\rangle \equiv (i \sigma_y |a\rangle)^*$, $\langle a|a'\rangle=0$ (note that $i\sigma_yi\sigma_y|a\rangle=-|a\rangle$). If the initial state is $|a\rangle$, this developer brings the system to the excited subspace. The quasiparticle tunnels to the dot, we measure outcome "1" and the system is in the state of the opposite parity, $|B\rangle|0\rangle$. (Fig. \ref{fig:developing}). If the initial state is orthogonal, no excitation occurs, we measure output "0" and get to the state $|b\rangle|0\rangle$. We see that the developer can be used to measure the probability of $|a\rangle$ if the initial parity is known to be even, and the final state is known from the measurement result. However, the parity is generally unknown.

Let us see how the same developer works in the odd subspace. We apply Eq.~(\ref{eq:UeUo}) to obtain:
\begin{eqnarray}
\label{eq:pulse1o}
D^{o}&=&-|B'\rangle|0\rangle\langle a'|\langle1|- |b'\rangle|1\rangle\langle a|\langle1|-\\ \nonumber&-&|B\rangle|0\rangle\langle A'|\langle0|+|b\rangle|1\rangle\langle A|\langle0|.
\end{eqnarray}
We see that now the developer tries to distinguish between $|A\rangle$ and $|A'\rangle$, while the final states for the same output are opposite: $|b\rangle|0\rangle$ for "1" and $|B\rangle|0\rangle$ for "0". Thus, we do not know the final state if the parity is unknown, neither we know which state has been measured. 

However, the situation can be fixed if we apply another unitary transformation. While this transformation does {\it not} depend on the result of the first measurement, it depends on the desired parity of the final state. In any case, the incoming states of a fixer are the same as the output states of the developer in the Majorana subspace. Let us consider the even fixer $F_{e}$ first. Its representation for two parities reads: 
\begin{eqnarray}
F_{e}^{e}&=&|c\rangle|0\rangle\langle b|\langle 0|+|C\rangle|1\rangle\langle b'|\langle 0|+\\ \nonumber&+&|C'\rangle|1\rangle\langle B|\langle1|+|c'\rangle|0\rangle\langle B'|\langle1|,\\
F_{e}^{o}&=&|c'\rangle|1\rangle\langle b'|\langle 1|+|C'\rangle|0\rangle\langle b|\langle 1|-\\ \nonumber&-&|C\rangle|0\rangle\langle B'|\langle0|-|c\rangle|1\rangle\langle B|\langle0|.
\end{eqnarray}
After the fixer, and the second measurement, the final state is always $|c\rangle|0\rangle$, this solves the initialization task. If the outcomes of the first and second measurements are "11" or "00", the initial parity was even. Otherwise, it was odd. 

The odd fixer $F_{o}$ has a similar structure,
\begin{eqnarray}
\label{eq:pulse2e1}
F_{o}^{e}&=&|C\rangle|1\rangle\langle b|\langle 0|+|c\rangle|0\rangle\langle b'|\langle 0|+\\ \nonumber&+&|c'\rangle|0\rangle\langle B|\langle1|+|C'\rangle|1\rangle\langle B'|\langle1|,\\
F_{o}^{o}&=&-|C'\rangle|0\rangle\langle b'|\langle 1|-|c'\rangle|1\rangle\langle b|\langle 1|+\\ \nonumber&+&|c\rangle|1\rangle\langle B'|\langle0|+|C\rangle|0\rangle\langle B|\langle0|.
\end{eqnarray}
In any case, the final state is $|C\rangle|0\rangle$. The measurement outcomes "11" and "00" manifest even initial parity, "01" and "10" manifest odd initial parity. So both fixers not only solve the initialization task: they determine the initial parity.

We see that the protocol described at the same time provides a measurement tool. Suppose we are able to arrange an unknown state of unknown parity, and reproduce it on demand. To characterize the state, one just repeats the protocol collecting the statistics of outcomes. 
The probabilities of outcomes "11","00","10","01" give the probabilities 
of the basis states $|a\rangle,|a'\rangle,|A\rangle,|A'\rangle$, respectively. The developer and fixer pulse sequences can be designed and realized for any choice of the superpositions $|a\rangle, |b\rangle, |c\rangle, |A\rangle, |B\rangle, |C\rangle$. In Supplemental Material \cite{supplemental}, we provide the concrete choice example.

\begin{figure}
	\centerline{\includegraphics[width=0.48\textwidth]{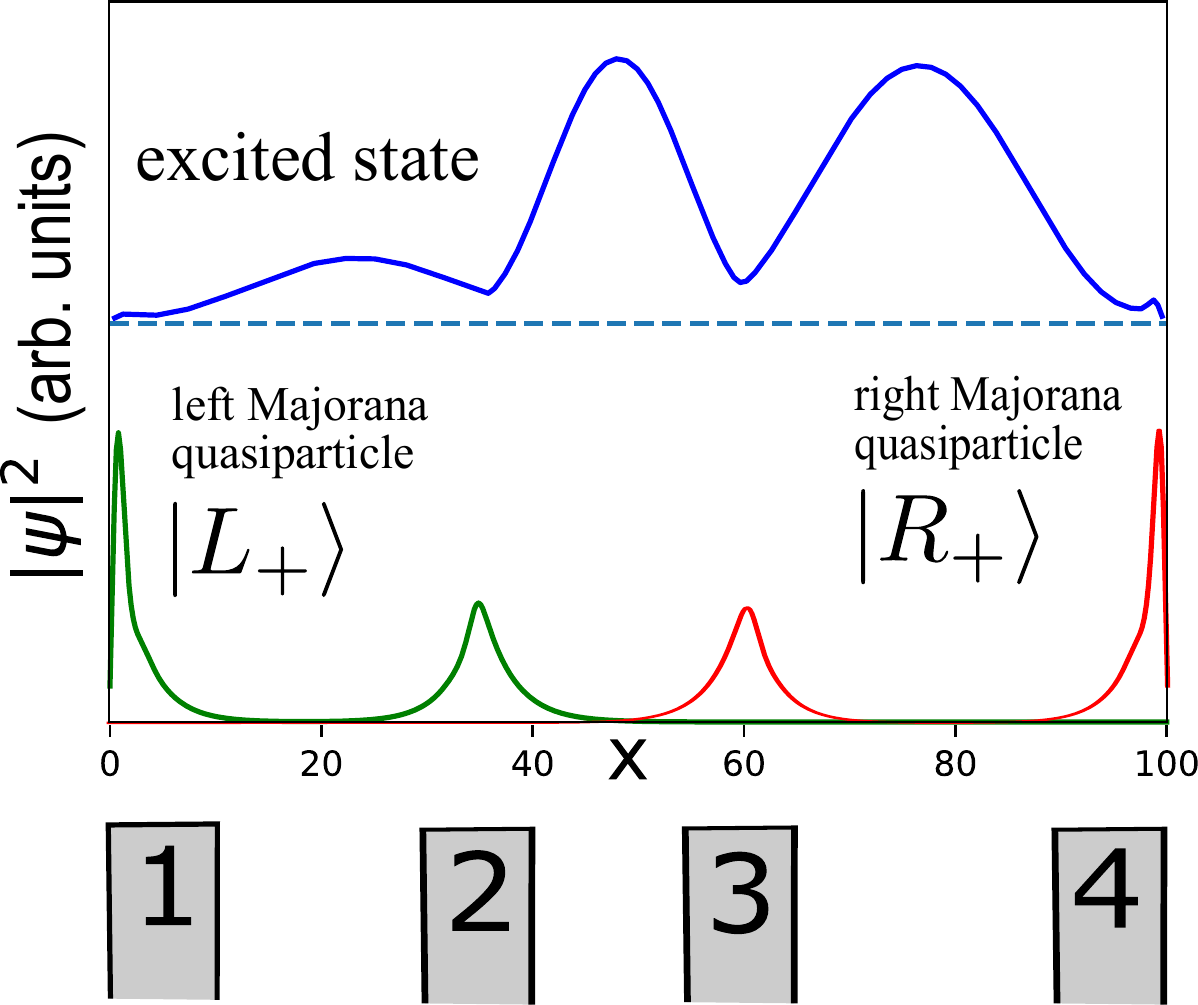}
	}
	\caption{The concrete illustrative setup. The probability densities of eigenfunctions and the positions of modulation gates $1-4$ chosen at $[0;10]$, $[30;40]$, $[55;65]$ and $[90;100]$.}
	\label{wfs}
\end{figure}
\begin{figure}
	\centerline{\includegraphics[width=0.48\textwidth]{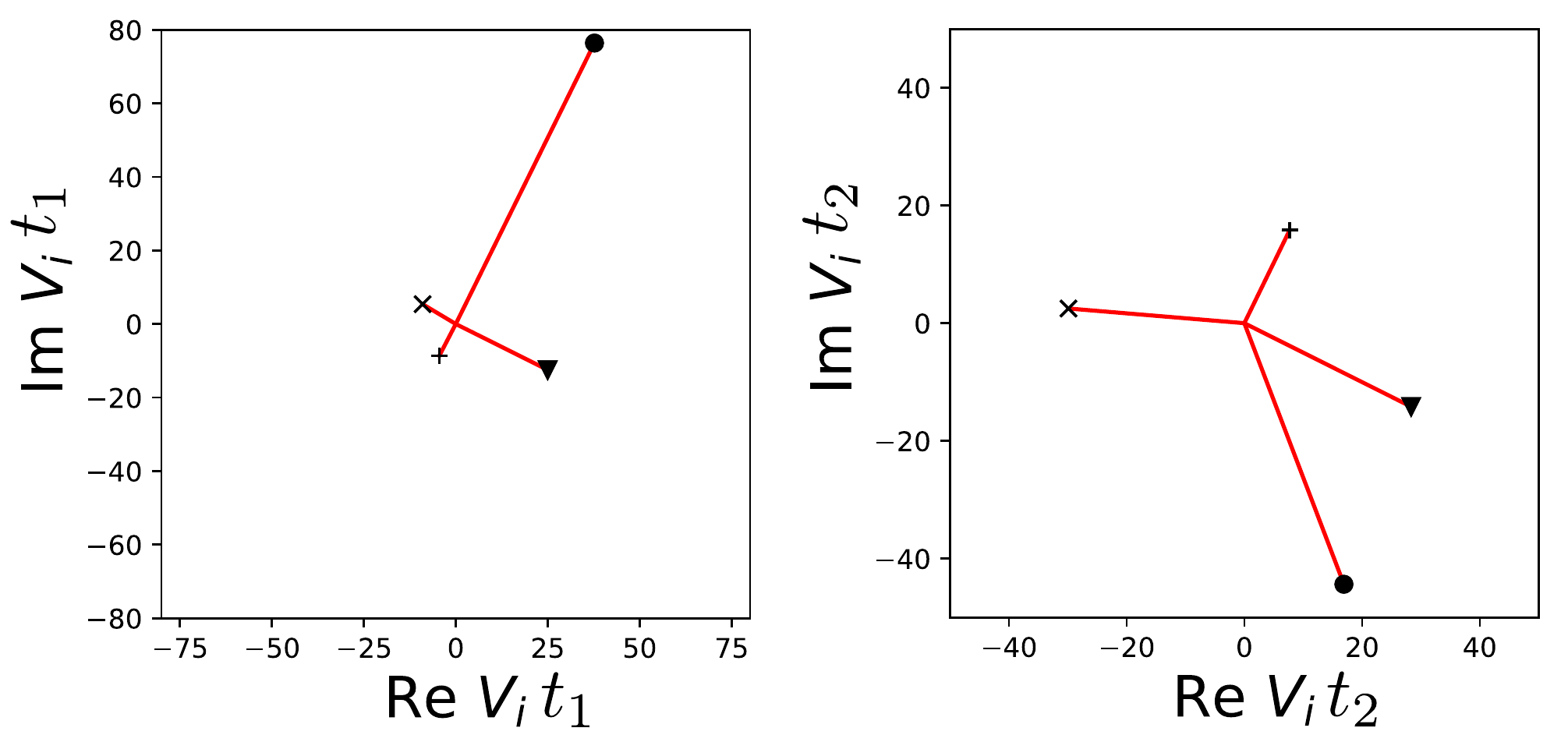}
	}
	\caption{Designed two-pulse sequence for the braiding $U_{23}$. The gate voltage amplitudes $V_{1-4}$ times pulse durations $t_{1,2}$ are given by circle, plus sign, triangle and x-mark, respectively.}
	\label{diagr}
\end{figure}

To show the feasibility of the setup and the suggested pulse sequence design, we now specify a microscopic model and provide extensive numerical study for a concrete set of parameters. 
We make use of the Hamiltonian \cite{lutchyn:prl10, oreg:prl10} to model a semiconducting nanowire with spin-orbit spectrum splitting, in the presence of applied uniform magnetic field $B$, and proximity-induced superconducting gap $\Delta$. The gap inverter gate is described by a coordinate-dependent potential $\mu(x)$ such that its values in the middle and outer sections, $\mu_{m},\mu_{o}$ satisfy the conditions of trivial  $B<\sqrt{\Delta^2+\mu^2_{m}}$ and non-trivial $B>\sqrt{\Delta^2+\mu^2_{o}}$ topology. The modulation gates are described by a time-dependent addition $\mu(x,t) = \sum_{i} V_i(t) \Theta(x-x_{i})\Theta(y_i-x)$, $x_i,y_i$ giving the start and end position of the gate $i$ (see Fig. \ref{wfs}).
The Hamiltonian in use reads
\begin{eqnarray}
\label{eq:H0}
&&H_0=\int dx\Psi^\dagger(x)\left[\left(-\frac{1}{2m}\frac{\partial^2}{\partial x^2} -i\alpha_{SOI}\sigma_z\frac{\partial}{\partial x}-\mu(x)\right)\tau_z\right.\nonumber \\ &&\left. +B\sigma_x+\Delta \tau_x\right]\Psi(x),\\
&&\Psi(x)=\{\psi_\uparrow(x), \psi_\downarrow(x),\psi_\downarrow^\dagger(x),-\psi_\uparrow^\dagger(x)\}, .\nonumber
\end{eqnarray}  
 $\psi_\sigma(x)$ being the electron field operators.  

We measure length and energy in units of $(m\alpha_{SOI})^{-1}$ and $m \alpha_{SOI}^2$, respectively. 
We compute the spectrum and wavefunctions diagonalizing the discrete-in-space appoximation of the Hamiltonian (\ref{eq:H0}), with the discretization step $0.2$. We choose a relatively long wire with length $L=100$ and the material parameters are of the order of 1: $B=3$, $|\Delta|=2.5$, $\mu_{m}=-1.91,\mu_{o}=-1.34$, see \cite{supplemental} for details. The transition between these two values are smoothed at the lenght scale of $3$, and the setup has been made slightly asymmetric. The bulk energy gaps correspondig to these parameters are $G_{e}=0.146$ and $G_{m}=0.164$, they are not precisely equal because of the finite size of the middle section. With this, the lowest excited state at $E_1 = 0.175$ is extended over the wire( see Fig. \ref{wfs}). Higher excited states are situated at $E_2=0.180 $ and $E_3=0.187$. For the resonant signal to address the lowest excited state only, the inverse pulse duration should not exceed the level spacing $E_2 - E_1$, this gives $t > 10^3$. 

The wave functions are presented in Fig. \ref{wfs}. There are 4 Majorana localized modes with the width $\approx 5$. We neglect a marginal overlap between the states setting them at zero energy. The wave function of the first excited state reminds the first particle-in-the box state with noticeable dips owing to orthogonality with Majorana peaks, and is extended over the whole length of the wire. With these wave functions, we compute the matrix elements of voltages applied to 4 gates whose positions are given in Fig. \ref{wfs}. This gives as a $4 \times 4$ matrix $\hat{M}$ that relates the voltage amplitudes and the resonant manipulation coefficients $\alpha_i$ (Eq. (\ref{eq:Hrm})). 
To design a pulse sequence corresponding to a unitary operation, we compute the resulting matrix depending on the parameters $\alpha_i$ and time duration of each pulse, and iteratively minimize in $\alpha_i$ the distance between the resulting and target matrix. Using the matrix $\hat{M}$, we convert to the gate voltage amplitudes. The design for the braiding of the second and the third Majorana mode is presented in Fig. \ref{diagr}, extensive examples are to be found in \cite{supplemental}.
  
To conclude, we propose a scheme that allows to realize braiding and all other unitary operations, as well as the measurement and initialization, for a Majorana qubit in a single 1D wire. It suits ideally to demonstrate macroscopically long coherence in Majorana space. The topological protection fails only during the operation. We illustrate the scheme with a concrete elaborated example. 

Let us shortly present necessary discussions in a wider context. No experimental system can be modelled with the accuracy we did. However, to design the pulse sequencies, one only needs $E_1$ and the matrix $M$: the latter can be determined from the analysis of the spectra of the dressed resonant state at varying $V_i$. The resonance with the lowest state only is essential since it minimizes dissipation. Moreover, the excitation to many excited states is exponentially suppressed owing to destructive interference. The scheme can be readily extended to {\it more} Majorana modes within the single wire, like proposed in \cite{dassarma:prb12, WimmerReview2020}.
While this can be done with a single state extended over the wire, but a simpler design would involve separate excited states, each extended over a group of Majorana modes. This can be achieved by proper profile of $\mu(x)$.  At the moment, the technological efforts are aimed to increase transparency of the barrier between the wire and the superconductor. As it is shown, for instance, in \cite{stanescu:prb11} at sufficiently high transparency the wire is not described by the Hamiltonian \cite{lutchyn:prl10, oreg:prl10} and eventually looses the localized excited states. So the moderate transparency is required for experimental realization of our idea. The idea presented may be also useful in the context of more traditional 2D Majorana braiding: one can set a localized excited state, switch on a resonant field, and move the modes passing the state to achieve the resonant manipulation and read out. 

\begin{acknowledgments} 
We acknowledge useful discussions with Anton Akhmerov, Kim P\"oyh\"onen and Felix von Oppen. This project has received funding from the European
Research Council (ERC) under the European Union's
Horizon 2020 research and innovation programme (grant
agreement No. 694272) and was supported by the Netherlands Organisation for
Scientific Research (NWO/OCW), as part of the Frontiers of Nanoscience (NanoFront)
program. The data that support the findings of this study
are available in \cite{data}.\end{acknowledgments}

\bibliographystyle{apsrev4-1}
\bibliography{majoranas_unitary_operations_bibliography}

\newpage
\renewcommand{\theequation}{S\arabic{equation}}
\setcounter{equation}{0}
\renewcommand{\thefigure}{S\arabic{figure}}
\renewcommand{\figurename}{Supplementary Fig.}

\setcounter{figure}{0}
\renewcommand{\thesection}{S\arabic{section}}
\setcounter{section}{0}

\begin{widetext}
\section*{Supplemental Material}
In this Supplemental Material, we present additional details and calculations regarding the example setup under consideration, as well as concrete designs of unitary transformations for quantum manipulation, initialization, and measurement. 

\section{S1. Wave functions and matrix elements of gate voltages \label{Sec:App0}}
To find the wave functions, we diagonalize numerically the Hamiltonian (\ref{eq:H0}). 
Owing to BdG symmetry, they come in pairs with positive and negative energies.
We fix the phases of these wavefunctions in such a way that $|\psi^*(E)\rangle=\sigma_x|\psi(E)\rangle$ and $|\psi(-E)\rangle=-i\tau_y \sigma_z |\psi(E)\rangle$.

This suffices for the wave functions of the excited state, $|{\rm ex}_{\pm}\rangle$,
$\pm$ corresponding to positive/negative energy. More work is required for wave functions in Majorana subspace. Owing to a residual overlap of Majorana modes (see Section S5), the eigenfunctions of the Hamiltonian are rather arbitrary linear combinations of the wave functions corresponding to the modes. To establish a proper basis in the Majorana subspace, we proceed as follows. We take 4 Hamiltonian eigenfunctions with lowest (residual) energies, form a matrix of elements of the operator $x$ in this 4-dimensional basis, and diagonalize it. The 4 eigenvalues correspond to 4 positions of the localized modes, and the corresponding eigenvectors are those of the modes. Next, we pick up two modes (1 and 2) on the left, and diagonalize $2 \times 2$ matrix of the elements of an operator $\tau_z$ (any operator with BdG symmetry would suffice). As the result, we obtain two eigenfunctions $|L_\pm\rangle$ of the Majorana quasiparticle on the left. Picking up two modes on the right (3 and 4), we construct $|R_{\pm}\rangle$.

Next, we compute the matrix elements of the perturbation $H_{\rm rm}$ brought by the gate voltage modulations, 
\begin{equation}
H_{\rm rm} = - \tau_z \sum_i V_i(t) \Theta(x-x_i)\Theta(y-y_i),
\end{equation} 
$x_i,y_i$ being start and end positions of the gate $i$.
We rewrite $H_{\rm rm}$ in the second-quantization form (Eq. \ref{eq:Hrm}) to the BdG form which allows us to express the coefficients $\alpha_i$ in terms of eigenstates of the BdG Hamiltonian with positive ($+$) and negative ($-$) energy eigenvalue, and find the following relations for the matrix elements,
\begin{align}
\label{eq:matrelem}
\alpha_1=\langle L_{-}|H_{\rm rm}|{\rm ex}_{+}\rangle; \;
\alpha_2=\langle R_{-}|H_{\rm rm}|{\rm ex}_{+}\rangle; \;
\alpha_3=\langle L_{+}|H_{\rm rm}|{\rm ex}_{+}\rangle; \;
\alpha_4=\langle R_{+}|H_{\rm rm}|{\rm ex}_{+}\rangle.
\end{align}

\section{S2. Relation between pulse parameters and gate voltage amplitudes \label{Sec:App1}}
The Hamiltonian for each resonant pulse is written in terms of resonant parameters $\alpha_{1-4}$. In odd subspace, it reads:
\begin{equation}
H=\begin{pmatrix}
0 & 0 & \alpha_4 & \alpha_1 \\
0 & 0 & \alpha_3 & -\alpha_2 \\
\alpha_4^* & \alpha_3^* & 0 & 0 \\
\alpha_1^* & -\alpha_2^* & 0 & 0 \\
\end{pmatrix}
\label{puls}
\end{equation}
Its form in even subspace is obtained  from the relation $H_{e} = -\Sigma_y \sigma_y H^*_{o} \sigma_y\Sigma_y$.

To find the relation between the resonant parameters and the gate voltage amplitudes, 
we evaluate Eqs. (\ref{eq:matrelem}) for each $V_i$ independently,
and invert the corresponding matrix. We obtain the linear relation $V_i = M_{ij} \alpha_j$ 
where the $4 \times 4$ real matrix $M$ is given by
\begin{equation}
\label{eq:matrixM}
M=\begin{pmatrix}
27.9300 &  -0.0009&-27.8803&0.0009\\
-5.8723 &  -0.0219&-5.8375&0.0219\\
0.0083  & 4.6067&0.0082&-4.5970\\
0.0002  &  -13.8756&0.0002 &-13.8682
\end{pmatrix}
\end{equation}

\section{S3. Pulse sequences required for braiding \label{Sec:App2}}
For 4 Majorana modes, there are six possible braiding matrices $U_{ij}$. We list here the explicit form of these matrices in the odd subspace: 

\begin{equation}
U^o_{13}=\frac{1}{\sqrt{2}}\begin{pmatrix}
1 & -1 & 0 & 0\\
1 & 1 & 0 & 0\\
0 & 0 & 1 & -1 \\
0 & 0 & 1 & 1
\end{pmatrix}
\label{13o}
\end{equation}
\begin{equation}
U^o_{12}=\frac{1}{\sqrt{2}}\begin{pmatrix}
1+i & 0 & 0 & 0\\
0 & 1-i & 0 & 0\\
0 & 0 & 1+i & 0 \\
0 & 0 & 0 & 1-i
\end{pmatrix}
\label{12o}
\end{equation}
\begin{equation}
U^o_{23}=\frac{1}{\sqrt{2}}\begin{pmatrix}
1 & i & 0 & 0\\
i & 1 & 0 & 0\\
0 & 0 & 1 & i \\
0 & 0 & i & 1
\end{pmatrix}
\label{23o}
\end{equation}
\begin{equation}
U^o_{14}=\frac{1}{\sqrt{2}}\begin{pmatrix}
1 & -i & 0 & 0\\
-i & 1 & 0 & 0\\
0 & 0 & 1 & i \\
0 & 0 & i & 1
\end{pmatrix}
\label{14o}
\end{equation}
\begin{equation}
U^o_{24}=\frac{1}{\sqrt{2}}\begin{pmatrix}
1 & -1 & 0 & 0\\
1 & 1 & 0 & 0\\
0 & 0 & 1 & 1 \\
0 & 0 & -1 & 1
\end{pmatrix}
\label{24o}
\end{equation}
\begin{equation}
U^o_{34}=\frac{1}{\sqrt{2}}\begin{pmatrix}
1-i & 0 & 0 & 0\\
0 & 1+i & 0 & 0\\
0 & 0 & 1+i & 0 \\
0 & 0 & 0 & 1-i
\end{pmatrix}
\label{34o}
\end{equation}
As mentioned, their form in the even subspace is obtained by transformation (\ref{eq:UeUo}) given in the main text.

To design the corresponding pulse sequence for a given target matrix $U$, we consider 2 pulses of resonant field with the Hamiltonians given by  \eqref{puls},
\begin{equation}
\Pi_2=e^{-iH_2 t_2}e^{-iH_1 t_2}
\end{equation}
$\Pi_2$ being the resulting matrix. We concentrate on the odd subspace. The resulting matrix depends on 8 complex parameters $\alpha^{(j)}_i t_j$, $j=1,2$, $t_j$ being the pulse durations. We define a distance in the space 
of unitary matrices,
\begin{equation}
D = {\rm Tr} \left((U-\Pi_2)(U-\Pi_2)^\dagger\right).
\end{equation}
We mininize $D$ iteratively in the space of $\alpha^{(j)}_i t_j$ starting a random initial point. If the minimum is achieved at $D=0$, we have the solution. If $D \ne 0$ at the minimum, we repeat the procedure.

For all braiding matrices (\ref{13o}), (\ref{12o}), (\ref{23o}), (\ref{14o}), (\ref{24o}), (\ref{34o}) we obtain the required parameters $\alpha^{(j)}_i t_j$ with the relative accuracy $\sim 10^{-3}$. Using the matrix $M$ given by Eq. (\ref{eq:matrixM}) we obtain the corresponding voltage amplitudes for each pulse. The results for all braiding matrices are collected in the Table \ref{table1}.
\begin{center}
	\begin{table}
		\resizebox{.2\textheight}{!}{%

			\begin{tabular}{ |c|c|c| } 
				\hline
				\multicolumn{3}{|c|}{$U^o_{13}$} \\
				\hline
				\multirow{4}{4em}{\rm first pulse} & 3.5626+1.4379i  & 104.8326 -3.3360i\\ 
				& 6.2875-0.7429i &  -51.6591 +8.0068i\\ 
				& 2.5388-0.6246i & -17.5693 +0.4805i \\ 
				& -0.2746+1.5416i & -45.5943-41.3344i\\ 
				\hline
				\multirow{4}{4em}{\rm second pulse} & -0.2407+1.3738i  & -79.6332 +2.4531i\\ 
				& -2.0651+1.3088i& 7.5805-14.8275i\\ 
				& 0.7874+1.2231i& -10.5582 +0.3552i\\ 
				& -2.5298+1.4434i& 38.4411-39.0808i\\ 
				\hline
				\multicolumn{3}{|c|}{$U^o_{12}$} \\
				\hline
				\multirow{4}{4em}{\rm first pulse} & -1.8112+0.9887i  & -38.0629-103.3642i\\ 
				& 3.1254-3.2631i & -44.6421 +16.6691i\\ 
				& 4.4961+0.4385i & 8.8887 -14.0691i \\ 
				& 0.1083-2.0622i & 23.6168 +14.9035i\\ 
				\hline
				\multirow{4}{4em}{\rm second pulse} & 1.2309-3.2858i  & 72.2241+133.8673i\\ 
				& 0.4281+1.5560i& 10.0026  +9.7060i\\ 
				& -2.1617-3.2428i& 21.4980 +18.1660i\\ 
				& 3.4413+0.6674i& -30.6798i +36.3059i\\ 
				\hline
				\multicolumn{3}{|c|}{$U^o_{23}$} \\
				\hline
				\multirow{4}{4em}{\rm first pulse} & -2.3878+1.1627i  & 37.7004+76.3880i \\ 
				& 1.0442+2.1060i & -4.4629 -8.6305i\\ 
				& -0.3063-0.6299i & 24.9729-12.4763i \\ 
				& 3.0368-1.5506i & -9.0232 +5.3913i\\ 
				\hline
				\multirow{4}{4em}{\rm second pulse} & -2.0021+1.4490i  & 16.9146-44.3686i\\ 
				& -0.3673-2.1380i& 7.7136+15.8356i\\ 
				& -0.9749-0.5503i& 28.3286-14.1843i\\ 
				& 4.1538-1.6281i& -29.8720 +2.4944i\\ 
				\hline

				\multicolumn{3}{|c|}{$U^o_{14}$} \\
				\hline
				\multirow{4}{4em}{\rm first pulse} & 0.6530+2.2008i  & -127.1431-100.2741i\\ 
				& -2.3252-3.1458i & 0.7380 +15.8719i\\ 
				& 2.2310+0.4450i & -22.5744  +0.7148i \\ 
				& -4.2483+2.3562i & 49.8913 -63.2159i\\ 
				\hline
				\multirow{4}{4em}{\rm second pulse} & -3.5774-2.0921i  & 102.6599-51.9062i\\ 
				& 0.6735-0.5148i& 13.4862 -4.8453i\\ 
				& -3.0076+1.3460i& 24.2822 +2.3053i\\ 
				& 1.7053+1.5887i& 25.9498+51.0587i\\ 
				\hline
				\multicolumn{3}{|c|}{$U^o_{24}$} \\
				\hline
				\multirow{4}{4em}{\rm first pulse} & -2.3317+0.5118i  &-69.7238-53.0538i
				\\ 
				& 0.6100+0.1589i & -21.7916-12.9138i\\ 
				& 3.1119+2.0622i & 9.0500-12.1150i  \\ 
				& -0.3689-2.1230i & 37.4569+22.3599i\\ 
				\hline
				\multirow{4}{4em}{\rm second pulse} & 4.4745+2.1958i & 144.9705+13.2449i\\ 
				& 2.2551+0.0303i& 3.9829 +2.4324i\\ 
				& -2.9404-0.4445i& -12.7219 -3.2558i\\ 
				& 1.7047+1.4852i& -85.7087-51.0616i\\ 
				\hline
				\multicolumn{3}{|c|}{$U^o_{34}$} \\
				\hline
				\multirow{4}{4em}{\rm first pulse} & -0.5448-1.7311i  & -138.1478-32.2069i\\ 
				& -1.6033-4.3352i & -10.1538+43.9911i\\ 
				& 3.3487-3.1879i & 4.3630+16.0604i\\ 
				& 0.4002+1.7723i & 2.0034 -0.5854i\\ 
				\hline
				\multirow{4}{4em}{\rm second pulse} & -0.4689-0.3569j & -41.7639+44.2386i\\ 
				& -1.9545-0.3445j& 14.1960+13.3048i\\ 
				& -0.4600-1.9318i& -6.8107 -0.8697i\\ 
				& -1.9420-0.5408i& 33.4493+12.4548i\\ 
				\hline
			\end{tabular}}
			\caption{Matrix elements of resonant perturbation for all braiding transformations. Left: from top to bottom $\alpha_4,\alpha_1,\alpha_3,\alpha_2$. Right: from top to bottom $V_1, V_2, V_3, V_4$}
			\label{table1}
		\end{table}
	\end{center}
\section{S4. Developer and fixer for initialization and measurement \label{Sec:App3}}
As discussed in the main text, we need two pulse sequences for measurement and initialization: a developer $D$ and a fixer $F$. For this illustration, we choose an even fixer $F_{e}$. In odd subspace, the corresponding unitary matrices are given by $\eqref{eq:pulse1o}$ and $\eqref{eq:pulse2e1}$. Applying two pulses brings the Majorana subsystem to $|c\rangle$.
As mentioned, the eigenstates involved can be chosen in arbitrary way.
The concrete choice we made for this example is as follows: 
\begin{equation}
|a\rangle=\begin{pmatrix}
1\\
0
\end{pmatrix},\quad |a^\prime\rangle=\begin{pmatrix}
0\\
-1
\end{pmatrix},\quad |b\rangle=\frac{1}{\sqrt{2}}\begin{pmatrix}
1\\
-i
\end{pmatrix},\quad|\beta^\prime\rangle=\frac{1}{\sqrt{2}}\begin{pmatrix}
i\\
-1
\end{pmatrix},\quad|c\rangle=\frac{1}{2}\begin{pmatrix}
\sqrt{3}\\
1
\end{pmatrix},\quad|c^\prime\rangle=\frac{1}{2}\begin{pmatrix}
1\\
-\sqrt{3}
\end{pmatrix}
\end{equation}
\begin{equation}
|B\rangle=\frac{1}{\sqrt{2}}\begin{pmatrix}
1\\
1
\end{pmatrix},\quad |B^\prime\rangle=\frac{1}{\sqrt{2}}\begin{pmatrix}
1\\
-1
\end{pmatrix},\quad |A\rangle=\begin{pmatrix}
i\\
0
\end{pmatrix},\quad|A^\prime\rangle=\begin{pmatrix}
0\\
i
\end{pmatrix},\quad|C\rangle=\frac{1}{3}\begin{pmatrix}
1\\
2\sqrt{2}i
\end{pmatrix},\quad|C^\prime\rangle=\frac{1}{3}\begin{pmatrix}
-2\sqrt{2}i\\
-1
\end{pmatrix}
\end{equation}

To design the corresponding pulse sequence, we apply the same numerical method as above. A peculiarity that the convergence for two pulses is rather poor. So for these transformations we implement three-pulse design:
\begin{equation}
\Pi_3=e^{-iH_3 t_3}e^{-iH_2 t_2}e^{-iH_1 t_3} .
\end{equation}
to achieve the relative accuracy $\sim 10^{-3}$. The resulting matrix elements $\alpha_i^{(j)} t_j$ and corresponding voltage amplitudes for these three pulses are presented in Table $\ref{table2}$.
\begin{center}
\begin{table}
		\begin{tabular}{ |c|c|c| } 
			\hline
			\multicolumn{3}{|c|}{$D_o$} \\
			\hline
			\multirow{4}{4em}{\rm first pulse} & -2.0495+2.0042i  & -61.1507+49.7052i\\ 
			& -4.0966+4.5633i & 35.1354-43.0467i\\ 
			& -1.9107+2.7887i & 15.8924 -6.3745i \\ 
			& 1.4154+0.6030i & 8.7827-36.1612i\\ 
			\hline
			\multirow{4}{4em}{\rm second pulse} & 6.5586-1.2382i  & 48.9201i-67.7397i\\ 
			& -1.3629-5.4106i& 26.3007+49.2372i\\ 
			& -3.1199-2.9906i& -17.6294 -1.3374i\\ 
			& 2.7259-1.5108i& -128.7829+38.1339i\\ 
			\hline
			\multirow{4}{4em}{\rm third pulse} & -1.4867-0.6612i& -16.1835-22.9873i\\ 
			& 0.5468i-0.1334i& -9.7742 -3.3047i\\ 
			& 1.1283+0.6907i & -4.9286+11.7841i\\ 
			& -2.5564+1.8971i& 56.0909-17.1536i\\ 
			\hline
			\multicolumn{3}{|c|}{$F_e^o$} \\
			\hline
			\multirow{4}{4em}{\rm first pulse} & 0.6126-0.8434i & -37.9861+72.2040i\\ 
			& -1.8684 +0.2338i & 13.9809+12.3891i\\ 
			& -0.5092-2.3555i  & -7.5534 -2.4738i\\ 
			& -1.0239-1.3748i & 5.7111+30.7735i\\ 
			\hline
			\multirow{4}{4em}{\rm second pulse}  & -0.3808+2.1101i & 39.7697-72.3703i\\ 
			& 0.0977 -0.4129i & 7.1449-10.2913i\\ 
			& -1.3286+2.1820i & 7.7986 -4.5144i\\ 
			& 1.3150+1.1225i & -12.9653-44.8405i\\ 
			\hline
			\multirow{4}{4em}{\rm third pulse} & 0.3121-1.2087i & 16.1518+13.2206i\\ 
			& -0.9123+0.4286i & 14.0443 -2.2930i\\ 
			& -1.4933-0.0448i & 6.4710 +7.8246i\\ 
			& 1.7204+0.4916i & -28.2014 +9.9404i\\ 
			\hline
		\end{tabular}
		\caption{Matrix elements of resonant perturbation for 2 necessary unitary transformations in parity measurement scheme. Left: from top to bottom $\alpha_4,\alpha_1,\alpha_3,\alpha_2$. Right: from top to bottom $V_1, V_2, V_3, V_4$}
		\label{table2}
\end{table}
\end{center}

\section{S5. Residulal overlap of Majorana modes}\label{Sec:App4}
For the example setup under consideration, we still have a remaining overlap between the separated Majorana modes, since their width $\simeq 5$ is only a factor of $5$ smaller than the minimal distance. Owing to the overlap, the two lowest eigenergies are finite, $4.14e-03$, and $2.33402015e-05$. If we express this as a Hamiltonian in the left-right basis in use, it reads
\begin{equation}
\begin{pmatrix}
0.00049355&0.00205414&-0.00205413&0\\
0.00205414&0.00019598&0&0.002054132\\
-0.00205413&0&-0.00019598&-0.002054143\\
0&0.002054132&-0.002054143&-0.00049355\\
\end{pmatrix}\begin{pmatrix}
L_{-}\\
R_{-}\\
L_{+}\\
R_{+}
\end{pmatrix}
\end{equation}
This Hamiltonian, in principle, results in unitary evolution at time scale $\simeq 1000$. We did not take this evolution into account neglecting the corresponding Hamiltonian. The overlap can be easily made exponentially smaller for longer setups. 

\end{widetext}

\end{document}